\documentclass{emulateapj}
\usepackage{epsf}

\begin{document}

%\shortauthors{Robertson \& Leiter}
%\shorttitle{Magnetic Moments in BHC}
\lefthead{Robertson \& Leiter}
\righthead{Magnetic Moments in BHC}

\title{On Intrinsic Magnetic Moments in Black Hole Candidates}

\author{Stanley L. Robertson\altaffilmark{1} and Darryl J. Leiter\altaffilmark{2}}
\altaffiltext{1}{Physics Dept.,Southwestern Oklahoma State University,
Weatherford, OK 73096 (roberts@swosu.edu)}
\altaffiltext{2}{FSTC, Charlottesville, VA 22901 (dleiter@aol.com)}

\begin{abstract}
In previous work we found that many of the spectral properties of low mass
x-ray binaries, including galactic black hole candidates
could be explained by a magnetic propeller model that requires an
intrinsically magnetized central object. Here we describe how the Einstein
field equations of General Relativity and equipartition magnetic fields
permit the existence of highly red shifted, extremely long lived, collapsing,
radiating objects. We examine the properties of these collapsed
objects and discuss characteristics that might lead to their confirmation
as the source of black hole candidate phenomena.
\end{abstract}


\section{Introduction}
In earlier work (Robertson \& Leiter 2002) we extended analyses of magnetic
propeller effects (Campana et al. 1998, Zhang, Yu \& Zhang 1998)
of neutron stars (NS) in low mass x-ray binaries (LMXB) to the domain
of galactic black hole candidates (GBHC). From the luminosities at the low/high
spectral state transitions, accurate rates of spin were found for NS
and accurate quiescent luminosities were calculated for both NS and GBHC.
NS magnetic moments were in agreement with those
found for similarly spinning 200 - 600 Hz pulsars.
GBHC spins were found to be typically 10 - 50 Hz. Their magnetic moments of
$\sim 10^{29}$ gauss cm$^3$ are $\sim 100$ times larger than those
of `atoll' class NS.
In the magnetic propeller model, the inner disk radius, $r$,
determines the spectral state. Very low to quiescent states correspond
to an inner accretion disk radius outside the light cylinder.
The inner disk radius lies between light cylinder and co-rotation radius
in the low/hard/radio-loud/jet-producing state of
the active propeller regime. The high/soft state corresponds to an
inner disk inside the co-rotation radius and accreting matter
impinging on the central object. We show here that this permits a
quantitative accounting for the `ultrasoft' high state spectral peak and
a high state hard x-ray spectral tail.

A field in excess of $10^8$ G has been found at the
base of the jets of GRS 1915+105 (Gliozzi, Bodo \& Ghisellini 1999,
Vadawale, Rao \& Chakrabarti 2001). A recent study of optical polarization
of Cygnus X-1 in its low state (Gnedin et al. 2003)
has found a slow GBHC spin and a magnetic field of $\sim 10^8$
gauss at the location of its optical
emission. Given the $r^{-3}$ dependence of field strength on magnetic
moment, the implied magnetic moments
are in good agreement with those we have found. 
Although Gnedin et al. attempted to explain the Cygnus X-1 magnetic field
as a result of a spinning charged black hole, the necessary charge
of $5\times 10^{28}$ esu would not be stable. Given the charge/mass ratios of
electrons and protons, the opposing electric forces on
them would then be at least $10^6$
times the gravitational attraction of $\sim 10 M_\odot$. Due to
highly variable accretion rates, it is also unlikely that
disk dynamos could produce the stability of fields needed to account for either
spectral state switches or quiescent spin-down luminosities.
Both also require magnetic fields co-rotating with the central object.

Considering the magnetic moments to be intrinsic to
the central object permits a physically obvious and unified
explanation of LMXB radio and spectral states,
but this is incompatible with the event horizons of black hole models 
of the GBHC. The success of the magnetic propeller model for GBHC
and the lack of evidence for event horizons in GBHC (Abramowicz,
Kluzniak \& Lasota 2002)
strongly suggests that it must be possible, within the confines of
Einstein's General Relativity to accommodate intrinsic magnetic
moments in gravitationally collapsed objects. This can be achieved
if the energy momentum tensor on the right hand side of the Einstein equation
\begin{equation}
G^{\mu\nu}=(8\pi G/c^4) T^{\mu\nu}
\end{equation}
is chosen in a manner that dynamically
enforces the Strong Principle of Equivalence (SPOE) requirement of 
`timelike worldline completeness'; i.e., the requirement that the worldlines
of physical matter, under the influence of both gravitational and 
non-gravitational forces, must remain timelike in all of spacetime 
(Wheeler \& Ciuofolini 1995). When this SPOE condition is met, trapped
surfaces leading to event horizons cannot be dynamically formed and 
intrinsic magnetic moments can exist in gravitationally collapsing 
objects (Leiter \& Robertson 2003, Mitra 2000, 2002, see below).

\section{Magnetospheric, Eternally Collapsing Objects (MECO)}
A relatively simple example of a collapsing, compact object that
can dynamically obey the SPOE requirement of
`timelike worldline completeness' is that of a radiating plasma containing an
equipartition magnetic dipole field that drives it to radiate at
its Eddington limit. Such an object can be described to first order
by the energy-momentum tensor:
\begin{equation}
T_{\mu}^{\nu} = (\rho + P/c^2)u_\mu u^\nu - P \delta_\mu^\nu + E_\mu^\nu
\end{equation}
where $E_\mu^\nu = qk_\mu k^\nu$, $k_\mu k^\mu = 0$ describes outgoing
radiation in a geometric optics approximation, $\rho$ is
energy density of matter, $P$ is the pressure and $q$ the flux of photon
radiation. For the collapsing mass, we use a comoving interior metric given by
\begin{equation}
ds^2=A(r,t)^2c^2dt^2 - B(r,t)^2dr^2- R(r,t)^2(d\theta^2 + sin^2\theta d\phi^2)
\end{equation}
and a non-singular exterior Vaidya metric with outgoing radiation
\begin{equation}
ds^2=(1-2GM/c^2R)c^2du^2 + 2c du dR - R^2(d\theta^2 + sin^2\theta d\phi^2)
\end{equation}
where $R$ is the areal radius and $u=t-R/c$ is the retarded observer time.
In order to maintain timelike worldline completeness as required by the 
SPOE, the surface redshift must remain finite (Leiter
\& Robertson 2003, Mitra 2000, 2002).
Then the proper time $d\tau_s$, at the collapsing,
radiating surface, S, will be positive definite if
\begin{equation}
d\tau_s= \frac{du}{1+z_s} = du(\Gamma_s+ U_s/c)~ >~ 0
\end{equation}
where $U_s = dR/d\tau$ is the proper time rate of change of $R(r,t)$ and
\begin{equation}
\Gamma_s =(1-\frac{2GM(r,t)_s}{c^2R_s} +\frac{U_s}{c})^{1/2}
\end{equation}
with M(r,t) the mass enclosed by the collapsing surface.

From Equations (5) and (6) we see that in order to satisfy the requirement of
timelike world line completeness for a collapsing object,
for which $U_s < 0$, it is necessary
to dynamically enforce the `no trapped surface condition',
$\frac{2GM_s}{c^2R_s} < 1$. In the MECO model, this is accomplished
by the non-gravitational force of outflowing radiation.
At the comoving surface, the luminosity is
$L_s = 4 \pi R^2 q~~ > 0$, where
\begin{equation}
q=-\frac{(c^2dM/d\tau)_s}{4 \pi R^2(\Gamma_s + U_s/c)}=\frac{L_\infty(1+z_s)^2}{4 \pi R^2}
\end{equation}
and the distantly observed luminosity is $L_\infty$.

\textit{To guarantee the existence of sufficient internal radiation pressure,
it is likely that a MECO must possess an equipartition magnetic dipole field.}
At the temperatures and compactness of stellar collapse, a pair plasma
exists within such a field. In addition to the intrinsic resistance to
collapse of magnetic flux (Thorne 1965), it has been shown
(Pelletier \& Markowith 1998) that the energy of magnetic perturbations in
equipartition pair plasmas is preferentially expended in photon production
rather than causing particle acceleration. Photon pressure
varies $\propto B^4$, due to its dependence on pair density ($\propto B^2$)
and synchrotron photon energy ($\propto B^2$). Lacking the pair plasma,
the ratio of magnetic ($\propto B^2$) to gravitational stresses would
be constant in a collapsing gas (e.g. Baumgarte \&
Shapiro 2003). With photon pressure capable of
increasing more rapidly than gravitational stress, a secular equilibrium rate
of collapse can be stabilized with the radiation temperature buffered near
the pair production threshold. The stability
of the rate of collapse is maintained by increased (decreased) photon
pressure ($\propto B^4$) if the field is increased (decreased) by compression
(expansion). An equipartition field also easily confines the pair plasma.
Thus the collapse differs in a fundamental way from that of only
weakly magnetic, radiation dominated polytropic gas or pressureless dust.

Strong recent evidence for equipartition magnetic fields in stellar collapse has
been found for GRB021206 (Coburn \& Boggs 2003) and strong residual fields much in
excess of those expected from mere flux compression
have been found in magnetars (Ibrahim, Swank \& Parke 2003).
Kluzniak and Ruderman (1998) have described the generation of $\sim 10^{17}$ G
magnetic fields for nuclear densities via differential
rotation in neutron stars. Other possibilities for producing extreme
magnetic fields would include ferromagnetic phase transitions during
the collapse (Haensel \& Bonnazzola 1996)
or the formation of quark condensates (Tatsumi 2000.)

Since distantly observed magnetic fields are reduced by
$\sim 1+z$, a redshift of $z \sim 10^8$ would be needed for the MECO
model with an equipartition field to accord with
the magnetic moments we have found for GBHC, and
also to account for AGN luminosity constraints (see the calculation
for Sgr A$^*$ below). \textit{Thus we are motivated by the SPOE
and empirical observational constraints to look for
solutions of the GR field equations that are consistent with objects in
extremely redshifted, Eddington limited gravitational collapse.}

\section{Eddington limited MECO}
The two key proper time differential equations
that control the behavior of the surface of an Eddington balanced, collapsing,
radiating object are:
(Hernandez Jr. \& Misner 1966, Lindquist, Schwartz \&
Misner 1965, Misner 1965):
\begin{equation}
\frac{dU_s}{d\tau} = (\frac{\Gamma}{\rho+P/c^2})_s (-\frac{\partial P}{\partial R})_s
- (\frac{GM}{R^2})_s
\end{equation}
Where $M_s = (M + 4\pi R^3(P + q )/c^2)_s$ includes
magnetic field energy in $P$ and radiant energy in $q$ and
\begin{equation}
\frac{dM_s}{d\tau}  =  - (4\pi R^2 P c \frac{U}{c})_s - (L (\frac{U}{c} + \Gamma))_s
\end{equation}
In Eddington limited steady collapse, the condition, $dU_s/d\tau \approx 0$, holds.
With this condition, Equation (8), when integrated over the
closed surface where the pressure is dominantly that of radiation, can be
solved for the net outward flow of Eddington limit luminosity through the
surface. Taking the escape cone factor of $27(GM_s/c^2R_s)^2/(1+z_s)^2$ into
account, the outflowing
(but not all escaping) surface luminosity, L, would be
\begin{equation}
L_{Edd}(outflow)_s  =\frac{4\pi G M_s c R^2(1 + z_{Edd,s})^3}
    {27 \kappa R_g^2}
\end{equation}
where $R_g=GM_s/c^2$ and $\kappa$ is the
plasma opacity. For simplicity, we have assumed here that the
luminosity actually escapes from the MECO surface rather than after
conveyance through a MECO pair photosphere.
The end result is the same for distant observers.
However the luminosity $L_s$ that appears in Equations (8 - 9) is actually
the net luminosity, which escapes through the photon sphere, and is given by
$L_{Edd}(escape)_s = L_{Edd}(outflow)_s - L_{Edd}(fallback)_s =
L_{Edd}(outflow)_s-L_{Edd}(outflow)_s(1-27R_g^2/(R(1+z_{Edd}))^2$
Thus in Equations (8) and (9), the $L_s$ appearing there is given by
\begin{equation}
L_s= L_{Edd}(escape)_s = \frac{4\pi GM(\tau)_s c (1+z_{Edd,s})}{\kappa}
\end{equation}
Due to the thermal buffering provided by the equipartition field and
pair plasma we can examine a limiting case
for which MECO mean proper density varies slowly enough that the condition
$U_s/c << 1/(1+z_s) \approx \Gamma_s$ also holds after a time, $\tau_{Edd}$,
that has elapsed in reaching the Eddington limited state. In this context
from (9) we have that
\begin{equation}
\frac{c^2dM_s}{d\tau} = -\frac{L_{Edd}(escape)_s}{1+z_s}
       =  - \frac{4\pi G M(\tau)_s c}{\kappa}
\end{equation}
which can be integrated to give
\begin{equation}
M_s(\tau) = M_s(\tau_{Edd}) \exp{((-4\pi G / \kappa c)(\tau - \tau_{Edd}))}
\end{equation}
For example, for hydrogen opacity, $\kappa = 0.4$ cm$^2$/g, and $z = 10^8$, this
yields a distantly observed MECO lifetime of
$(1+z_s)\kappa c/4\pi G \sim 5\times 10^{16}$ yr.
The MECO state for GBHC is likely preceded by a much faster
gravitational collapse of a stellar core. With a neutrino opacity
some $10^{20}$ times smaller than that of photons,
the lifetime of Eddington limited
neutrino emissions would likely be minutes, at most.
To stabilize the rate of collapse with magnetic
pressure and synchrotron generated photons would require a photon luminosity
reduced below the neutrino Eddington luminosity
by the same factor of $\sim 10^{20}$,
to a distantly observed $\sim 10^{32}$ erg/s. It is of interest to note that
for this to correspond to a MECO object radiating at its local Eddington limit,
a surface redshift of $z_s \sim 10^{7-8}$ would be required, which accords
with our earlier arguments based on empirical magnetic field observational
constraints. 

\section{The quiescent MECO}
Distantly observed MECO luminosity is diminished by
$1/(1+z)^2$ by gravitational redshift and by
$27 R_g^2/(R(1+z))^2 \sim 27/(4(1+z)^2)$ by a narrow escape cone at
the photosphere of a pair atmosphere. Due to its
negligible mass, we consider the pair atmosphere to be external to the
MECO. It can be shown \textit{ex post facto} to be radiation dominated such
that $T^4/(1+z)$ is constant throughout. Estimates of luminosities,
photosphere upper limit temperatures and photosphere redshifts can
then be found from the Eddington balance requirements.

The fraction of luminosity from the MECO surface
that escapes to infinity in Eddington balance is
\begin{equation}
(L_{Edd})_s = \frac {4\pi G M_s c(1+z)}{\kappa} = 1.27 \times 10^{38}m(1+z_s)~~erg/s
\end{equation}
where $m = M/M_\odot$. The distantly observed luminosity is:
\begin{equation}
L_\infty = \frac{(L_{Edd})_s}{(1+z_s)^2} = \frac {4\pi G M_s c}{\kappa(1+z_s)}
=\frac{1.27 \times 10^{38}m}{(1+z_s)}~~erg/s
\end{equation}
By assuming that the escaping radiation is primarily thermal and that
the photosphere temperature is $T_p$,
the fraction that escapes to be distantly observed is:
\begin{equation}
L_\infty = \frac{4 \pi R_g^2 \sigma T_p^4 27}{(1+z_p)^4}
    = 1.56\times 10^7 m^2 T_p^4 \frac{27}{(1+z_p)^4}~~erg/s
\end{equation}
where $\sigma = 5.67\times 10^{-5}$ erg/s/cm$^2$
and subscript p refers to conditions at the photosphere.
Equations (15) and (16) yield:
\begin{equation}
T_\infty = T_p/(1+z_p) = \frac{2.3\times 10^7}{(m(1+z_s))^{1/4}}~~ K.
\end{equation}
To examine typical cases, a GBHC with $m = 10$
and $z \sim 10^8$ would have
$T_\infty = 1.3\times 10^5 K = 0.01$ keV, a luminosity,
excluding spin-down contributions, of
$L_\infty =1.3\times 10^{31} erg/s$, and a spectral peak at 220 A$^0$,
in the photoelectrically absorbed deep UV.
For an m=$10^8$ AGN, $T_\infty = 2300 K$, and
$L_\infty = 1.3\times 10^{38} erg/s$
with a spectral peak in the near infrared at 1.2 micron.
\textit{(Sgr A$^*$, with $m=3\times 10^6$, would have $T_\infty=5500$ K and
a 2.2 micron brightness of 6 mJy, just below the observational upper limit
of 9 mJy (Reid et al. 2003).)}
Since $T_\infty = T_p/(1+z_p)$, $T_p^4/(1+z_p) = T_s^4/(1+z_z)$
 and $T_s \approx 6\times10^9$ K, we find that
\begin{equation}
T_p = T_s(\frac{T_s}{T_\infty (1+z_s)})^{1/3}
= 3.8\times 10^{10}\frac{m^{1/12}}{(1+z_s)^{1/4}} ~~K
\end{equation}
For a GBHC with $m=10$ and $z_s=10^8$, this yields a photosphere temperature
of $4.6\times 10^8$ K, from which $(1+z_p) = 3500$. An AGN with $m=10^8$
would have a somewhat warmer photosphere at $T_p = 1.8\times 10^9$ K, but
with a red shift of $7.7\times 10^5$.

Hence, although they are not black holes, passive MECO without accretion 
disks would (using any realistic opacity) have lifetimes much greater 
than a Hubble time and emit highly red
shifted quiescent thermal spectra that may be quite
difficult to observe. 

\section{The High State of An Actively Accreting MECO}
From the viewpoint of a distant observer, accretion would
deliver mass-energy to the MECO, which would then radiate most of it
away. The contribution from the central MECO alone would be
\begin{equation}
L_\infty = \frac {4\pi G M_s c}{\kappa(1+z_s)}+ \frac{\dot{m}_\infty c^2}{1+z_s}(e(1+z_s)-1)
    = 4 \pi R_g^2 \sigma T_p^4 \frac {27} {(1+z_p)^4}
\end{equation}
where $e = E/m_0 c^2 = 0.943$ is the specific energy per particle
available after accretion disk flow to the marginally stable orbit radius,
$r_{ms}$. Assuming that $\dot{m}_\infty$ is some fraction, f, of the
Newtonian Eddington limit mass accretion rate, $4\pi G M c/\kappa$, then
\begin{equation}
1.27\times 10^{38}\frac{m\eta}{1+z_s} =
(27)(1.56\times 10^7)m^2(\frac{T_p}{1+z_p})^4
\end{equation}
where $\eta=1+f((1+z_s)e -1)$ includes both quiescent and accretion
contributions to the luminosity.
Due to the extremely strong dependence on temperature of the
density of pairs, it is likely that the photosphere temperature
remains near the previously found $4.6\times 10^8 K$.
Then with $z_s=10^8$, $m=10$, and $f=1$, we find
$T_\infty = T_p/(1+z_p) = 1.3\times 10^7$ K = 1.1 keV,
and $(1+z_p) = 35$, which indicates considerable photospheric expansion.
The MECO luminosity would be $L_\infty = 1.2\times 10^{39}$ erg/s,
which is approximately at the Newtonian Eddington limit. 
For comparison, the accretion disk outside the marginally stable orbit at $r_{ms}$
(efficiency = 0.057) would produce only $6.8\times 10^{37}$ erg/s, with
an inner disk temperature also `ultrasoft' at $\sim 1.1$ keV. 

Most photons escaping the photon sphere
would depart with some azimuthal momentum on spiral
trajectories that would eventually take them across and through the
accretion disk. Thus a very large fraction of the soft photons would be subject
to bulk comptonization in the plunging region inside $r_{ms}$.
This contrasts sharply with the situation for neutron stars
where there is no comparably large plunging region. This
accounts for the fact that hard x-ray spectral tails are
comparatively much stronger for high state GBHC.
Our preliminary calculations for photon trajectories
randomly directed upon leaving the
photon sphere indicate that this process
would produce a power law component with photon index greater than 2.

\section{Detecting MECO}
It may be possible to detect MECO in several ways. Firstly, for a red shift
of $z \sim 10^8$, the quiescent luminosity of a GBHC MECO would be
$\sim 10^{31} erg/s$ with $T_\infty \sim 0.01$ keV. This thermal peak
in the strongly absorbed UV might be
observable for very nearby  or high galactic latitude GBHC,
such as A0620-00 or XTE J1118+480. 
Secondly, At high state luminosities above $\sim 10^{36}$ erg/s,
a central MECO would be a bright, small `ultrasoft' central object that might
be sharply eclipsed in deep dipping sources. Thirdly, a pair plasma atmosphere
in an equipartition magnetic field should be virtually transparent to photon
polarizations perpendicular to the magnetic field lines. The x-rays
from the central MECO should exhibit some polarization that might be
detectable. If GBHC MECO are the offspring of massive
star supernovae, then they should be found all over the galaxy. Based
upon our estimates of their quiescent temperatures, isolated GBHC MECO
would be weak, polarized, EUV sources with a power-law tail in soft x-rays.

\end{document}